\documentstyle[11pt,aaspp4,flushrt]{article}

\newcommand{\cmjj}{\mbox{${\rm cm^{-2}}$}}

\newcommand{\kms}{\mbox{km\ s${^{-1}}$}}
\newcommand{\lya}{\mbox{${\rm Ly}\alpha$}}

\begin{document}

\title{DAMPED \lya\ ABSORPTION ASSOCIATED WITH AN EARLY-TYPE GALAXY AT REDSHIFT
$z = 0.16377$\altaffilmark{1,2}}

\author{KENNETH M. LANZETTA\altaffilmark{3}, ARTHUR M. WOLFE\altaffilmark{4},
HAKAN ALTAN\altaffilmark{3}, \\
XAVIER BARCONS\altaffilmark{5,6}, HSIAO-WEN CHEN\altaffilmark{3}, ALBERTO
FERN\'{A}NDEZ-SOTO\altaffilmark{3}, \\
DAVID M. MEYER\altaffilmark{7}, AMELIA ORTIZ-GIL\altaffilmark{3}, SANDRA
SAVAGLIO\altaffilmark{8}, JOHN K. WEBB\altaffilmark{9}, \\
and NORIAKI YAHATA\altaffilmark{3}}

\altaffiltext{1}{Based on observations with the NASA/ESA Hubble Space
Telescope, obtained at the Space Telescope Science Institute, which is operated
by the Association of Universities for Research in Astronomy, Inc., under NASA
contract NAS5--26555.}

\altaffiltext{2}{Based on observations made with the William Herschel Telescope
operated on the island of La Palma by the Royal Greenwich Observatory in the
Spanish Observatorio del Roque de Los Muchachos of the Instituto de
Astrof\'{\i}sica de Canarias.}

\altaffiltext{3}{Astronomy Program, Department of Earth and Space Sciences,
State University of New York at Stony Brook, Stony Brook, NY 11794--2100,
U.S.A.}

\altaffiltext{4}{Center for Astrophysics and Space Sciences, University of
California, San Diego, La Jolla, CA 92093--0111, U.S.A.}

\altaffiltext{5}{Instituto de F{\'\i}sica de Cantabria (Consejo Superior de
Investigaciones Cient{\'\i}ficas---Universidad de Cantabria), Facultad de
Ciencias, 39005 Santander, SPAIN}

\altaffiltext{6}{Institute of Astronomy, Madingley Road, Cambridge CB3 0HA,
ENGLAND}

\altaffiltext{7}{Department of Physics and Astronomy, Northwestern University,
Evanston, IL 60208, U.S.A.}

\altaffiltext{8}{European Southern Observatory, K. Schwarzschild Str.\ 2,
Garching b.\ M\"unchen, D--85748, GERMANY}

\altaffiltext{9}{School of Physics, University of New South Wales, P.O. Box 1,
Kensington, NSW 2033, AUSTRALIA}

\newpage

\begin{abstract}

  We report new HST and ground-based observations of a damped \lya\ absorption
system toward the QSO 0850$+$4400.  The redshift of the absorption system is $z
= 0.163770 \pm 0.000054$ and the neutral hydrogen column density of the
absorption system is $\log N = 19.81 \pm 0.04$ \cmjj.  The absorption system is
by far the lowest redshift confirmed damped \lya\ absorption system yet
identified, which provides an unprecedented opportunity to examine the nature,
impact geometry, and kinematics of the absorbing galaxy in great detail.  The
observations indicate that the absorption system is remarkable in three
respects:  First, the absorption system is characterized by weak metal
absorption lines and a low metal abundance, possibly less than 4\% of the solar
metal abundance.  This cannot be explained as a consequence of obscuration by
dust, because the neutral hydrogen column density of the absorption system is
far too low for obscuration by dust to introduce any significant selection
effects.  Second, the absorption system is associated with a
moderate-luminosity early-type S0 galaxy, although the absorption may actually
arise in one of several very faint galaxies detected very close to the QSO line
of sight.  Third, the absorbing material moves counter to the rotating galaxy
disk, which rules out the possibility that the absorption arises in a thin or
thick co-rotating gaseous disk.  These results run contrary to the expectation
that low-redshift damped \lya\ absorption systems generally arise in the gas-
and metal-rich inner parts of late-type spiral galaxies.  We suggest instead
that mounting evidence indicates that low-redshift galaxies of a variety of
morphological types may contain significant quantities of low metal abundance
gas at large galactocentric distances.

\end{abstract}
 

\newpage

\section{INTRODUCTION}

  Damped \lya\ absorption systems detected toward background QSOs trace the
gaseous content of the universe to very high redshifts.  Most surveys for
damped \lya\ absorption systems have been carried out at optical wavelengths
(Wolfe et al.\ 1986; Lanzetta et al.\ 1991; Wolfe et al.\ 1995;
Storrie-Lombardi, Irwin, \& McMahon 1996), where the absorbers are detected at
redshifts ranging from $z \approx 1.6$ (the redshift of \lya\ at the
atmospheric cutoff) through $z \approx 5$ (the redshift of the most distant
background QSOs in the surveys).  These optical surveys have demonstrated that
damped \lya\ absorption systems dominate the mass density of neutral gas in the
universe, containing roughly enough gas at $z \approx 3.5$ to form the stars of
present-day galaxies.  But although the redshifts sampled by the optical
surveys span an important epoch in the history of galaxy evolution, the cosmic
time interval spanned by the optical surveys corresponds to only a small
fraction of the current age of the universe.

  To address this issue, we are conducting a spectroscopic survey of damped
\lya\ absorption systems at ultraviolet wavelengths, based on archival
International Ultraviolet Explorer (IUE) and Hubble Space Telescope (HST)
low-resolution spectroscopy and targeted HST high-resolution spectroscopy
(Lanzetta, Wolfe, \& Turnshek 1995; Lanzetta et al.\ in preparation).  A
primary goal of the survey is to identify damped \lya\ absorption systems at
very low redshifts, in order both to measure the gaseous content of the nearby
universe and to secure a sample of nearby absorbers for detailed study.  We
have so far surveyed ultraviolet spectra of more than 500 QSOs, BL Lac objects,
and Seyfert galaxies, and we have obtained confirming (or otherwise) spectra of
most of the candidate damped \lya\ absorption systems identified in
low-resolution or low signal-to-noise ratio observations.

  Full results of the survey will be presented elsewhere, but here we report
new HST and ground-based observations of a damped \lya\ absorption system
identified in the survey at redshift $z = 0.16377$ toward the QSO 0850$+$4400. 
The absorption system is of particular interest because it was identified not
only in our spectroscopic survey of damped \lya\ absorption systems but also in
our imaging and spectroscopic survey of galaxies and absorption systems toward
HST spectroscopic target QSOs (Lanzetta et al.\ 1995), in which it was found to
be associated with a galaxy just $16.6 \ h^{-1}$ kpc from the QSO line of
sight.  The absorption system is by far the lowest redshift confirmed damped
\lya\ absorption system yet identified, which provides an unprecedented
opportunity to examine the nature, impact geometry, and kinematics of the
absorbing galaxy in great detail.

  The HST observations reported here include (1) a new Goddard High Resolution
Spectrograph (GHRS) spectrum of 0850$+$4400 and (2) a new Wide Field Planetary
Camera 2 (WFPC2) image of the field surrounding 0850$+$4400.  The ground-based
observations reported here include (3) a new spectrum of the galaxy associated
with the absorption system.  A dimensionless Hubble constant $h = H_0/(100 \
{\rm km} \ {\rm s}^{-1} \ {\rm Mpc}^{-1})$ and a deceleration parameter $q_0 =
0.5$ are adopted throughout.

\section{OBSERVATIONS}

\subsection{GHRS Spectroscopy}

  Spectroscopic observations of 0850$+$4400 were obtained with HST using GHRS
with the G140L grating on 28 September 1996.  The observations were obtained in
a series of 16 exposures each of 300 s duration for a total exposure time of
4800 s.  The individual exposures were reduced using standard pipeline
techniques and were registered to a common origin and coadded using our own
reduction programs.  The final spectrum was fitted with a smooth continuum
using an iterative spline fitting technique.  The spectral resolution of the
final spectrum was measured to be ${\rm FWHM} = 0.60$ \AA, and the continuum
signal-to-noise ratio of the final spectrum was measured to be $S/N \approx 12$
per resolution element.  A portion of the spectrum is shown in Figure 1.

\subsection{WFPC2 Imaging}

  Imaging observations of objects in the field surrounding 0850$+$4400 were
obtained with HST using WFPC2 with the F702W filter on 7 February 1996.  The
observations were obtained in a series of three exposures each of 800 s
duration for a total exposure time of 2400 s.  The individual exposures were
reduced using standard pipeline techniques and were registered to a common
origin, filtered for cosmic rays, and coadded using our own reduction programs.
The spatial resolution of the final image was measured to be ${\rm FWHM}
\approx 0.1$ arcsec, and the $5 \sigma$ detection threshold of the final image
to point sources was measured to be $AB(7020) = 26.3$.  A portion of the image
is shown in Figure 2.

\subsection{WHT Spectroscopy}

  Spectroscopic observations of galaxy $-$00089$+$00020 were obtained with the
William Herschel Telescope (WHT) using the ISIS double spectrograph with two
1200 line mm$^{-1}$ gratings and two Tektronix CCD detectors on 4 March 1995. 
(Here and throughout galaxies are indicated by noting their coordinate offsets
in Right Ascension and Declination, respectively, from the QSO line of sight in
units of 0.1 arcsec.)  The observations were obtained in a series of three
exposures each of 1800 s duration for a total exposure time of 5400 s.  A long
slit of width 1.5 arcsec was aligned to a position angle of 50 deg, which is
roughly coincident with the major axis of the galaxy.  Observations of a Cu-Ar
and a Cu-Ne arc lamp were obtained following each exposure, and observations of
a tungsten lamp were obtained at the end of the night.  The individual
exposures were bias subtracted, flatfielded, and extracted using our own
reduction programs, and the resulting spectra were wavelength calibrated by
means of fifth-order polynomial fits to spectral lines identified in the arc
lamp observations. The RMS residuals of the polynomial fits were measured to be
0.03 \AA\ (of the blue spectrum) and 0.05 \AA\ (of the red spectrum).  The
wavelength scales were reduced to vacuum, heliocentric values.  The spectral
resolutions of the final spectra were measured to be ${\rm FWHM} = 1.14$ \AA\
(of the blue spectrum) and ${\rm FWHM} = 1.07$ \AA\ (of the red spectrum).

\subsection{Other Observations}

  In addition to the new observations described above, other observations are
relevant to the present analysis:  First, we obtained imaging and spectroscopic
observations of 16 faint galaxies and stars (including galaxy $-$00089$+$00020)
in the field surrounding 0850$+$4400 (Lanzetta et al.\ 1995; Lanzetta et al.\
in preparation).  Second, Bahcall et al.\ (1995) obtained Faint Object
Spectrograph (FOS) G160L, G190H, and G270H spectra of 0850$+$4400, which we
accessed through the HST archive.  These other observations will be considered
as appropriate in the analysis described below.

\section{ANALYSIS}

\subsection{Absorption System Spectroscopy}

  In this section we consider the GHRS G140L spectrum described in \S\ 2.1 and
the FOS G190H and G270H spectra mentioned in \S\ 2.4 in order to (1) measure
physical parameters of the absorption system and (2) characterize the metal
content of the absorption system.

  To measure physical parameters of the absorption system, we applied the
$\chi^2$ Voigt profile fitting routine described by Lanzetta \& Bowen (1992)
to the GHRS G140L spectrum, adopting the redshift, Doppler parameter, and
neutral hydrogen column density as parameters.  The resulting fit is shown in
Figure 1.  The redshift of the absorption system is
\begin{equation}
z = 0.163770 \pm 0.000054
\end{equation}
and the neutral hydrogen column density of the absorption system is
\begin{equation}
\log{N} = 19.81 \pm 0.04 \ \cmjj,
\end{equation}
which corresponds to a rest-frame equivalent width of
\begin{equation}
W = 5.9 \pm 0.3 \ {\rm \AA}.
\end{equation}
The Doppler parameter of the absorption system is essentially unconstrained
because the absorption line occurs on the damped part of the curve of growth.
To determine whether the total column density estimate could be affected by
unresolved velocity structure, we repeated the analysis allowing for two
absorbing components.  We found that it is not possible to obtain a fit with a
significantly lower total column density.  Regardless of how the initial
parameter values were set, the final solution always converged to a total
column density consistent with the value of equation (2) to within formal
uncertainties.

  To characterize the metal content of the absorption system, we searched for
prominent metal absorption lines at the redshift of the absorption system in
the GHRS G140L and FOS G190H and G270H spectra.  The GHRS G140L spectrum covers
predicted absorption lines of N V, O I, Si II, and S II; the FOS G190H spectrum
covers predicted absorption lines of C IV and Si IV; and the FOS G270H spectrum
covers predicted absorption lines of Mg II, Mn II, and Fe II.  The resulting
measurements or upper limits are given in Table 1.  Of 26 predicted absorption
lines covered by the spectra, absorption lines are tentatively detected only at
the predicted wavelengths of C IV $\lambda 1548$, Mg II $\lambda 2803$, Si II
$\lambda 1193$, and Si II $\lambda 1260$.  But no absorption lines are detected
at the predicted wavelengths of corresponding C IV $\lambda 1550$, Mg II
$\lambda 2796$, Si II $\lambda 1190$, or Si II $\lambda 1304$, suggesting that
the detected absorption lines might be unrelated chance coincidences.  The most
stringent limit to the metal abundance of the absorption system is set by the
absence of detectable O I $\lambda 1302$ absorption, because O$^0$ has an
ionization potential very nearly equal to one Rydberg and ${\rm O}^0 / {\rm
H}^0$ is expected to trace ${\rm O} / {\rm H}$ under a variety of ionization
conditions.  But the upper limit to the O$^0$ column density depends
sensitively on the Doppler parameter, which is unknown.  We therefore
calculated the column densities corresponding to the O I $\lambda 1302$
equivalent width limit of Table 1 for assumed Doppler parameters of 30, 10, 5,
and 2 \kms.  The resulting upper limits are given in Table 2.  For any Doppler
parameter satisfying $b > 10$ \kms, the $2 \sigma$ upper limit to the O$^0$
column density satisfies $\log N < 15.39$ \cmjj, which adopting the neutral
hydrogen column density of equation (2) corresponds to a $2 \sigma$ upper limit
to the O abundance of $[{\rm O}/{\rm H}] < -1.36$, or less than 4\% of the
solar metal abundance.  The O abundance could be appreciably higher than this
only if the Doppler parameter is less than 10 \kms, which cannot be ruled out
by the present observations.

\subsection{Galaxy Imaging}

  In this section we consider the WFPC2 F702W image described in \S\ 2.2 and
the imaging and spectroscopic observations of faint galaxies and stars
mentioned in \S\ 2.4 in order to (1) measure morphological and spectral
parameters of the galaxy and (2) search for faint galaxies very close to the
QSO line of sight.

  The absorption system is associated with galaxy $-$00089$+$00020, which the
imaging and spectroscopic observations of Lanzetta et al.\ (1995) indicate
occurs at redshift $z \approx 0.1634$ and impact parameter $\rho = 16.6 \
h^{-1}$ kpc.  These observations spectroscopically identify a total of 16 faint
galaxies and stars in the field but no other galaxies at redshift $z \approx
0.16$.

  To measure morphological parameters of the galaxy, we applied the $\chi^2$
disk plus bulge profile fitting routine described by Chen et al.\ (1997) to the
WFPC2 F702W image, adopting the position, disk and bulge effective (or
half-light) surface brightnesses, disk and bulge effective (or half-light)
radii, inclination angle and axis ratio, and orientation angle as parameters.
The disk effective radius of the galaxy is
\begin{equation}
R_D = 1.4 \pm 0.3 \ h^{-1} \ {\rm kpc},
\end{equation}
and the bulge effective radius of the galaxy is
\begin{equation}
R_B = 2.5 \pm 0.2 \ h^{-1} \ {\rm kpc}.
\end{equation}
The disk inclination angle of the galaxy is
\begin{equation}
i = 50 \pm 2 \ {\rm deg},
\end{equation}
and the position angle of the major axis of the galaxy is
\begin{equation}
{\rm PA} = 52 \pm 2 \ {\rm deg},
\end{equation}
which forms an angle to the projected line segment joining the galaxy to the
QSO of
\begin{equation}
\alpha = 51 \pm 2 \ {\rm deg}.
\end{equation}
Directly integrating the theoretical surface brightness profiles, the
integrated disk-to-bulge ratio of the galaxy is
\begin{equation}
D/B = 0.4 \pm 0.1,
\end{equation}
and the apparent magnitude of the galaxy is
\begin{equation}
AB(7020) = 19.21 \pm 0.01.
\end{equation}
To measure spectral parameters of the galaxy, we applied the galaxy spectral
classification routine described by Yahata et al.\ (in preparation) to the
spectrum of galaxy $-$00089$+$00020 of Lanzetta et al.\ (1995).  The galaxy is
spectrally classified as an Sb galaxy, and the rest-frame $B$-band absolute
magnitude of the galaxy (applying an appropriate $K$ correction calculated from
the spectral energy distributions of Coleman, Wu, \& Weedman 1980) is
\begin{equation}
M_B - 5 \log h = -18.61 \pm 0.01.
\end{equation}
Taking the rest-frame $B$-band absolute magnitude of an $L_*$ galaxy to be
$M_{B_*} - 5 \log h = -19.5$, this corresponds to rest-frame $B$-band
luminosity of
\begin{equation}
L_B = 0.4 L_{B_*}.
\end{equation}

  To search for faint galaxies very close to the QSO line of sight, we
subtracted a Tiny Tim (Krist 1995) point spread function model of the QSO from
the WFPC2 F702W image.  The resulting image is shown in Figure 3.  The image
reveals several very faint galaxies very close to the QSO line of sight, for
which spectroscopic identifications have not yet been obtained.  Galaxy
$+$00004$+$00019 is of angular separation 1.9 arcsec and apparent magnitude
$AB(7020) = 22.5$, galaxy $-$00026$+$00024 is of angular separation 3.5 arcsec
and apparent magnitude $AB(7020) = 24.5$, and galaxy $-$00023$+$00043 is of
angular separation 4.9 arcsec and apparent magnitude $AB(7020) = 26.1$.  If
these galaxies are at the redshift of the absorption system, then they occur at
impact parameters ranging from $\rho = 3.8 \ h^{-1}$ kpc (galaxy
$+$00004$+$00019) to $\rho = 8.7 \ h^{-1}$ kpc (galaxy $-$00023$+$00043) and
rest-frame $B$-band luminosities (applying the $K$ correction calculated for
galaxy $-$00089$+$00020) ranging from $L_B = 0.002 L_{B_*}$ (galaxy
$-$00023$+$00043) to $L_B = 0.05 L_{B_*}$ (galaxy $+$00004$+$00019).
Alternatively, the galaxies might occur at the redshift $z = 0.513$ of the QSO,
in which case they are not related to the absorption system.

\subsection{Galaxy Spectroscopy}

  In this section we consider the WHT ISIS spectrum described in \S\ 2.3 in
order to (1) measure spectral properties of galaxy $-$00089$+$00020 and (2)
compare the rotation curve of the galaxy with the velocity of the absorption
system.

  To measure spectral properties of the galaxy, we obtained Gaussian fits to
the H$\alpha$ and [N II] emission lines of the WHT ISIS spectrum, adopting the
systemic redshift, velocity width, and peak intensity as parameters.  The
systemic redshift of the galaxy is
\begin{equation}
z = 0.163483 \pm 0.000011,
\end{equation}
which adopting the redshift of the absorption system of equation (1)
corresponds to a velocity difference between the galaxy and the absorption
system of
\begin{equation}
74 \pm 14 \ \kms.
\end{equation}

  To compare the rotation curve of the galaxy with the velocity of the
absorption system, we extracted a spatially- and spectrally-resolved image of
the H$\alpha$ emission line of the WHT ISIS spectrum.  Following Barcons,
Lanzetta, \& Webb (1995), we corrected the projected velocity $v$ and impact
parameter $\rho$ to the line joining the center of the galaxy to the QSO line
of sight by assuming circular motion of a flat disk.  The resulting rotation
curve together with a point indicating the velocity and impact parameter
difference between the galaxy and the absorption system is shown in Figure 4.
The absorbing material does not participate in the rotation of the galaxy disk.
Rather the absorbing material moves counter to the rotating galaxy disk,
missing the predicted motion of the galaxy disk by more than 150 \kms\ at an
impact parameter of $16.6 \ h^{-1}$ kpc.  This observation rules out the
possibility that the absorbing material arises in a thin or thick co-rotating
gaseous disk.

\subsection{Nature of Galaxy $-$00089$+$00020}

  Considering the visual appearance, size, integrated disk-to-bulge ratio, and
rest-frame $B$-band luminosity, we conclude that galaxy $-$00089$+$00020 is a
moderate luminosity early-type S0 galaxy.

\section{SUMMARY AND DISCUSSION}

  The main results of the analysis are summarized as follows:

  1.  The damped \lya\ absorption system toward 0850$+$4400 identified in our
spectroscopic survey of damped \lya\ absorption systems at ultraviolet
wavelengths (Lanzetta, Wolfe, \& Turnshek 1995; Lanzetta et al.\ in
preparation) and in our imaging and spectroscopic survey of galaxies and
absorption systems toward HST spectroscopic target QSOs (Lanzetta et al.\ 1995)
is confirmed by a GHRS G140L spectrum.  The redshift of the absorption system
is $z = 0.163770 \pm 0.000054$ and the neutral hydrogen column density of the
absorption system is $\log N = 19.81 \pm 0.04$ \cmjj.  The absorption system is
by far the lowest redshift confirmed damped \lya\ absorption system yet
identified, which provides an unprecedented opportunity to examine the nature,
impact geometry, and kinematics of the absorbing galaxy in great detail.

  2.  The absorption system is characterized by weak metal absorption lines and
low metal abundances.  No absorption lines of N V, O I, Si IV, S II, Mn II, or
Fe II are detected to within sensitive upper limits, although weak absorption
lines of C IV, Mg II, and Si II may be present.  The most stringent limit to
the metal abundance of the absorption system is set by the absence of
detectable O I absorption, which places a $2 \sigma$ upper limit to the O
abundance of $[{\rm O}/{\rm H}] < -1.36$ (or less than 4\% of the solar O
abundance) if the Doppler parameter of the absorption system satisfies $b > 10$
\kms\ (although a smaller Doppler parameter and a correspondingly larger O
abundance cannot be ruled out by the present observations).

  3.  The absorption system is associated with a moderate-luminosity early-type
S0 galaxy.  The the impact parameter of the galaxy is $\rho = 16.6 \ h^{-1}$
kpc and the rest-frame $B$-band luminosity of the galaxy is $L_B = 0.4
L_{B_*}$.  No other galaxies in the field are identified at a comparable
redshift, and no other galaxies of luminosity exceeding $L_B = 0.05 L_{B_*}$
are present at smaller impact parameters.  The absorption may actually arise in
one of several very faint galaxies detected very close to the QSO line of sight
for which spectroscopic identifications have not yet been obtained.

  4.  The redshift of the galaxy is $z = 0.163483 \pm 0.000011$, and the
velocity difference between the galaxy and the absorption system is $74 \pm
14$ \kms.  The absorbing material moves counter to the rotating galaxy disk,
which rules out the possibility that the absorption arises in a thin or thick
co-rotating disk.

  Our first conclusion is that these results run contrary to the expectation
that low-redshift damped \lya\ absorption systems generally arise in the gas-
and metal-rich inner parts of late-type spiral galaxies, which are thought to
dominate the gaseous content of the nearby universe (Rao \& Briggs 1993). 
Instead these results fit a pattern that appears to be emerging from
observations of several other damped \lya\ absorption systems at redshifts $z <
1.6$.  Specifically, a number of recent observations indicate that low-redshift
damped \lya\ absorption systems very often (1) exhibit low metal abundances
(with evidence of Population II metal abundance patterns) (Meyer \& York 1992;
Steidel et al.\ 1995; Meyer, Lanzetta, \& Wolfe 1995) and (2) arise far from
the inner regions of galaxies of a variety of morphological types (Steidel et
al.\ 1997; Le Brun et al.\ 1997).  In contrast, there are {\em no} known
examples of low-redshift damped \lya\ absorption systems that exhibit solar or
near-solar metal abundances and only one known example of a low-redshift damped
\lya\ absorption system that arises in the inner region of late-type spiral
galaxy (Le Brun et al.\ 1997).  Why is it that apparently ``anomalous''
observations appear to be the rule rather than the exception?

  Previous authors have argued that obscuration by dust might preferentially
favor selection of chemically unevolved galaxies of lower dust content over
chemically evolved galaxies of higher dust content (e.g.\ Steidel et al.\ 1994;
Pei \& Fall 1995; Steidel et al.\ 1997).  But it is easy to show that the
neutral hydrogen column density of the absorption system toward 0850$+$4400 is
far too low for obscuration by dust to introduce any significant selection
effects.  Even adopting a Galactic dust-to-gas ratio and dust extinction curve,
the predicted $V$-band extinction caused by the absorption system amounts to no
more than 0.05 mag.  Yet the $V$-band magnitude of 0850$+$4400 is $V = 16.62$,
which is almost 1.5 mag brighter than the nominal magnitude limit $V \approx
18$ of the HST survey for damped \lya\ absorption systems.  Hence the
absorption system would have been included into the survey essentially
irrespective of dust content, and the observations {\em cannot} be explained as
a consequence of obscuration by dust.  Similar arguments indicate that
obscuration by dust is also unlikely to explain the observations of the other
low-redshift damped \lya\ absorption systems, which are also generally
characterized by low neutral hydrogen column densities (c.f.\ Meyer, Lanzetta,
\& Wolfe 1995).

  We suggest instead that mounting evidence indicates that the observations
must be taken seriously as indicative of the gaseous content of the
low-redshift universe.  Because damped \lya\ absorption systems dominate the
mass density of neutral gas in the low-redshift universe (Lanzetta, Wolfe, \&
Turnshek 1995), a straightforward interpretation is that the observations
indicate that galaxies of a variety of morphological types may contain
significant quantities of low metal abundance gas at large galactocentric
distances.  This gas could constitute roughly half of the mass density of
neutral gas in the low-redshift universe, based on the very limited statistics
that are so far available.  The existence of such gas can in principle be
confirmed by means of additional observations of low-redshift damped \lya\
absorption systems.

  Our second conclusion is that these results run contrary to the hypothesis
that damped \lya\ absorption systems arise in rotating ensembles of clouds
(Lanzetta \& Bowen 1992; Prochaska \& Wolfe 1997).  This hypothesis was
advanced to explain the observation that metal absorption lines of damped \lya\
absorption systems very often exhibit asymmetric absorption profiles that are
characteristic of rotational motion.  Of course, in this case no significant
metal absorption lines are detected, so it is not possible to tell whether the
absorption system does or does not exhibit this signature.  The asymmetric
absorption profiles have been previously identified only in connection with
higher column density absorbing material (i.e.\ with neutral hydrogen column
densities satisfying $\log N > 20.3$ \cmjj), hence it is possible that higher
column density absorption systems arise in rotating ensemble of clouds and
lower column density absorption systems do not.  Similar observations of other
low-redshift galaxy and absorption system pairs are needed to establish the
relationship between the stellar and gaseous kinematics of galaxies.

  The observations reported here have two other notable implications as
follows:

  First, the observation that the absorption system is characterized by weak
metal absorption lines (e.g. with no detectable Mg II $\lambda \lambda 2796,
2803$ doublet) demonstrates that low-redshift damped \lya\ absorption systems
cannot be reliably identified on the basis of metal absorption lines.  Previous
attempts to identify low-redshift damped \lya\ absorption systems from known Mg
II absorption systems (e.g.\ Rao, Turnshek, \& Briggs 1995) are apparently
biased toward absorption systems with stronger metal absorption lines and
higher metal abundances.

  Second, the observation of the reality of the absorption system helps to
establish the anti-correlation between \lya\ absorption equivalent width and
galaxy impact parameter found in our previous analysis (Lanzetta et al.\ 1995). 
Subsequent authors doubted the reality of the absorption system and excluded it
from consideration in their analyses, which failed to detect the
anti-correlation between \lya\ absorption equivalent width and galaxy impact
parameter (e.g.\ Le Brun, Bergeron, \& Boisse 1996).  The galaxy and absorption
system pair toward 0850$+$4400 (of \lya\ absorption equivalent width $W = 5.9$
\AA\ and galaxy impact parameter $\rho = 16.6 \ h^{-1}$ kpc) apparently must be
included into any objective and unbiased analysis of the relationship between
\lya\ absorption equivalent width and galaxy impact parameter.

\acknowledgements

  The authors thank the staff of STScI for their expert assistance.  H.A.,
H.-W.C., A.F-S., K.M.L., A.O.-G., and N.Y. were supported by grant NAG--53261
from NASA; grants AR--0580--30194A, GO--0594--80194A, GO--0594--90194A, and
GO--0661--20195A from STScI; and grant AST--9624216 from NSF.  X.B. was
partially supported by the DGES under project PB95--0122 and acknowledges
sabbatical support at Cambridge by the DGES under grant PR95---490.

\newpage

\begin{center}
\begin{tabular}{p{1.5in}cc}
\multicolumn{3}{c}{TABLE 1} \\
\multicolumn{3}{c}{REST-FRAME EQUIVALENT WIDTHS OF} \\
\multicolumn{3}{c}{METAL ABSORPTION LINES$^{\rm a}$}
\\
\hline
\hline
\multicolumn{1}{c}{Absorption Line} & $W$ (\AA) & $\sigma_W$ (\AA) \\
\hline
C IV  $\lambda 1548$ \dotfill & 0.23     & 0.06 \\
C IV  $\lambda 1550$ \dotfill & $< 0.04$ & ...  \\
N V   $\lambda 1238$ \dotfill & $< 0.07$ & ...  \\
N V   $\lambda 1242$ \dotfill & $< 0.08$ & ...  \\
O I   $\lambda 1302$ \dotfill & $< 0.09$ & ...  \\
Mg II $\lambda 2796$ \dotfill & $< 0.19$ & ...  \\
Mg II $\lambda 2803$ \dotfill & 0.43     & 0.11 \\
Al II $\lambda 1670$ \dotfill & $< 0.16$ & ...  \\
Si II $\lambda 1190$ \dotfill & $< 0.07$ & ...  \\
Si II $\lambda 1193$ \dotfill & 0.15     & 0.03 \\
Si II $\lambda 1260$ \dotfill & 0.10     & 0.03 \\
Si II $\lambda 1304$ \dotfill & $< 0.10$ & ...  \\
Si IV $\lambda 1393$ \dotfill & $< 0.63$ & ...  \\
Si IV $\lambda 1402$ \dotfill & $< 0.44$ & ...  \\
S II  $\lambda 1250$ \dotfill & $< 0.08$ & ...  \\
S II  $\lambda 1253$ \dotfill & $< 0.08$ & ...  \\
S II  $\lambda 1259$ \dotfill & $< 0.07$ & ...  \\
Mn II $\lambda 2576$ \dotfill & $< 0.19$ & ...  \\
Mn II $\lambda 2594$ \dotfill & $< 0.20$ & ...  \\
Mn II $\lambda 2606$ \dotfill & $< 0.20$ & ...  \\
Fe II $\lambda 1608$ \dotfill & $< 0.10$ & ...  \\
Fe II $\lambda 2344$ \dotfill & $< 0.18$ & ...  \\
Fe II $\lambda 2374$ \dotfill & $< 0.19$ & ...  \\
Fe II $\lambda 2382$ \dotfill & $< 0.19$ & ...  \\
Fe II $\lambda 2586$ \dotfill & $< 0.19$ & ...  \\
Fe II $\lambda 2600$ \dotfill & $< 0.19$ & ...  \\
\hline
\multicolumn{2}{l}{\hspace{0.25in} $^{\rm a}$Upper limits are $2 \sigma$.}
\end{tabular}
\end{center}

\newpage

\begin{center}
\begin{tabular}{p{1.5in}c}
\multicolumn{2}{c}{TABLE 2} \\
\multicolumn{2}{c}{UPPER LIMITS TO} \\
\multicolumn{2}{c}{${\rm O}^0$ COLUMN DENSITY$^{\rm a}$} \\
\hline
\hline
\multicolumn{1}{c}{$b$ (\kms)} & $\log N$ (\cmjj) \\
\hline
30 \dotfill & $< 14.56$ \\
10 \dotfill & $< 15.39$ \\
5  \dotfill & $< 17.87$ \\
2  \dotfill & $< 18.17$ \\
\hline
\multicolumn{2}{l}{\hspace{0.25in} $^{\rm a}$Upper limits are $2 \sigma$.}
\end{tabular}
\end{center}

\newpage

\figcaption{Portion of GHRS G140L spectrum of 0850$+$4400 centered on \lya\
absorption line.  Spectral resolution is ${\rm FWHM} = 0.60$ \AA\ (or ${\rm
FWHM} = 130$ \kms), and continuum signal-to-noise ratio is $S/N \approx 12$ per
resolution element.  Spectrum has been smoothed at roughly the Nyquist rate.
Smooth curve shows $\chi^2$ Voigt profile fit, which indicates a redshift $z =
0.163770 \pm 0.000054$ and a neutral hydrogen column density $\log N = 19.81
\pm 0.04$ \cmjj.}

\figcaption{Portion of WFPC2 F702W image of field surrounding 0850$+$4400. 
Spatial resolution is ${\rm FWHM} = 0.1$ arcsec, and angular extent is $40
\times 40$ arcsec$^2$.  Indicated objects have been spectroscopically
identified (Lanzetta et al.\ 1995; Lanzetta et al.\ in preparation).  Object
$-$00089$+$00020 is a galaxy of redshift $z = 0.1635$, object $+$00000$+$00000
is 0850$+$4400, object $-$00001$-$00050 is a star, object $+$00100$-$00024 is a
galaxy of redshift $z = 0.4402$, and object $+$00226$+$00124 is a galaxy of
redshift $z = 0.5007$.}

\figcaption{Portion of WFPC2 F702W image of field surrounding 0850$+$4400 for
which a Tiny Tim point spread function model of the QSO has been subtracted. 
Spatial resolution is ${\rm FWHM} = 0.1$ arcsec, and angular extent is $12
\times 12$ arcsec$^2$.  Indicated objects are at small angular separations to
the QSO.  Object $+$00004$+$00019 is a galaxy of angular separation 1.9 arcsec
and apparent magnitude $AB(7020) = 22.5$, object $-$00026$+$00024 is a galaxy
of angular separation 3.5 arcsec and apparent magnitude $AB(7020) = 24.5$, and
object $-$00023$+$00043 is a galaxy of angular separation 4.9 arcsec and
apparent magnitude $AB(7020) = 26.1$.}

\figcaption{Spatially- and spectrally-resolved image of H$\alpha$ emission line
of WHT ISIS spectrum of galaxy $-$00089$+$00020 together with point indicating
velocity and impact parameter difference between  galaxy and absorption system.
Spectral resolution is ${\rm FWHM} = 1.07$ \AA\ (or ${\rm FWHM} = 42$ \kms),
and spatial resolution is ${\rm FWHM} = 1.69$ arcsec (or ${\rm FWHM} = 3.01 \
h^{-1}$ kpc).  Following Barcons, Lanzetta, \& Webb (1995), the projected
velocity $v$ and impact parameter $\rho$ are corrected to the line joining the
center of the galaxy to the QSO line of sight by assuming circular motion of a
flat disk.  Contours are equally spaced, at arbitrary levels of intensity.}

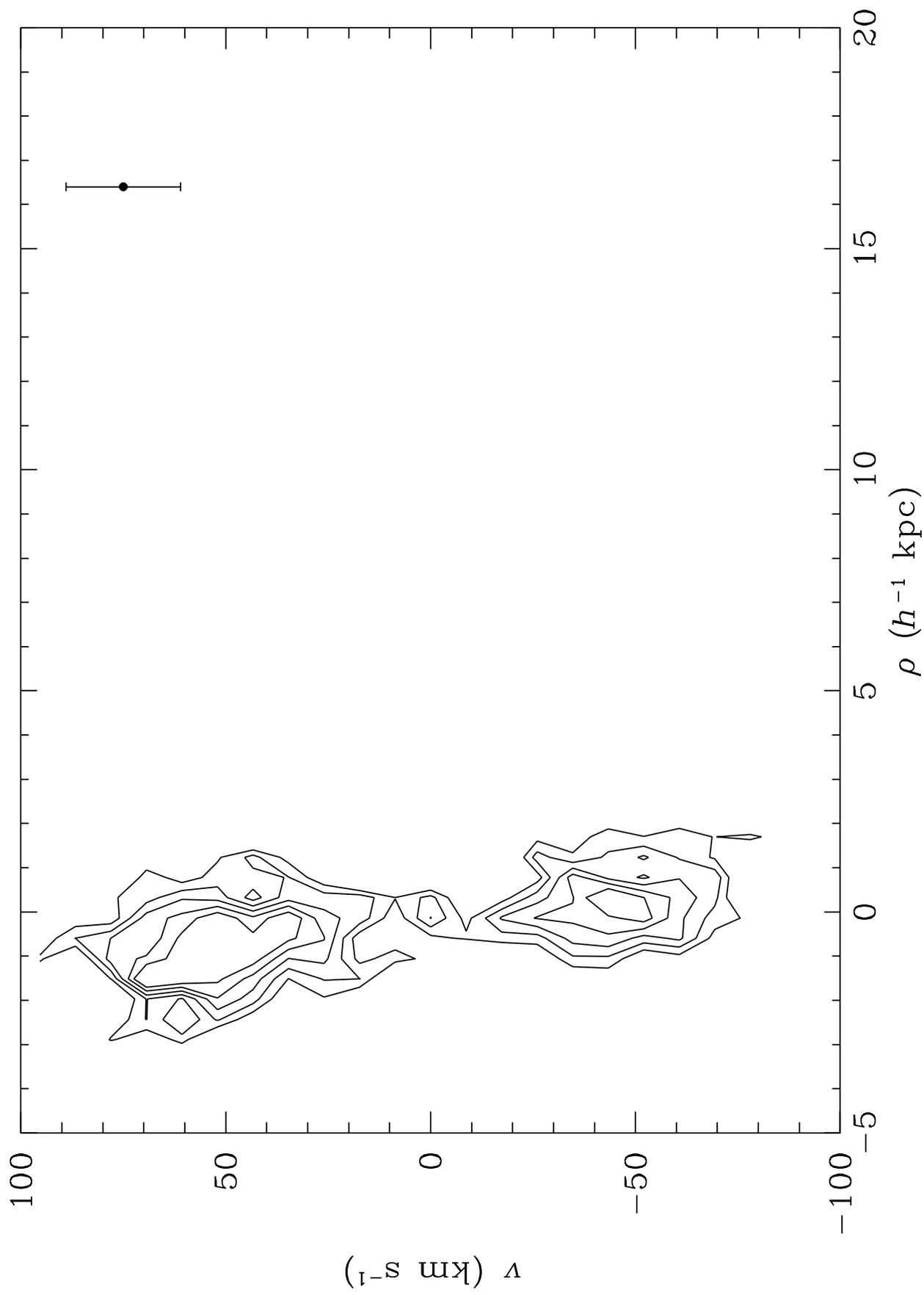
\end{document}